\documentclass{appolb}

\newcommand{\AmS}{{\protect\the\textfont2
  A\kern-.1667em\lower.5ex\hbox{M}\kern-.125emS}}

\begin{document}
\title{The Breaking of Isospin and the $\rho$--$\omega$--System}

\author{H. Fritzsch\address{Sektion Physik,
        Ludwig--Maximilians--Universit\"at M\"unchen,\\
        Theresienstrasse 37, D--80333 M\"unchen}
        \thanks{Supported by Deutsche Forschungsgemeinschaft,
		DFG-No. FR 412/25--2}}
\maketitle
\begin{abstract}
Simple quark models for the low lying vector mesons suggest a
mixing between the $u$-- and $d$--flavors and a violation of
the isospin symmetry for the $\rho-\omega$ system much stronger
than observed. It is shown that the chiral dynamics, especially
the QCD anomaly, is responsible for a restoration of the isospin
symmetry in the $\rho-\omega$ system.
\end{abstract}
\maketitle


Although there are no doubts that all observed strong interaction
phenomena can be described within the theory of QCD, a quantitative
description of the strong interaction phenomena in the low energy
sector is still lacking,
although some features of the low energy phenomena have been
partially understood by the lattice gauge theory approach.\\
The low energy sector of the physics of the strong interactions
is dominated by the low--lying pseudoscalar mesons
($\pi, K,\eta,\eta'$) and the low--lying vector mesons
($\rho,\omega, K^{*},\phi$). It is well--known that the structures of
the quark wave functions of the pseudoscalar mesons ($0^{-+}$) and of
the vector mesons ($1^{--}$) differ substantially.\\
In the vector meson channel there is a strong mixing between           
the eights component of the $SU(3)$ octet (wave function:
$(\bar u u +\bar d d -2\bar s s)/(\sqrt 6)$) and of the $SU(3)$
singlet (wave function: $(\bar u u +\bar d d +\bar s s)/(\sqrt 3)$).
The mixing strength is such that the mass eigenstates are nearly
the state $(\bar u u +\bar d d)/(\sqrt 2)$), the $\omega$--meson,
and the state $\bar s s$, the $\phi$--meson. While this feature looks
peculiar, when viewed upon from the platform of the underlying $SU(3)$
symmetry, it finds a simple interpretation, if one takes into account
the Zweig rule \cite{okub63}, which states that the
mixing must take place
in such a way that quark lines are neither destroyed nor created.\\
On the other hand the pseudoscalar mesons follow the pattern
prescribed by the $SU(3)$ symmetry in the absence of singlet--octet
mixing. The neutral mass eigenstates $\eta$ and $\eta'$ are nearly
an $SU(3)$--octet or $SU(3)$--singlet:
\begin{eqnarray}
\quad\eta &\approx& \frac{1}{\sqrt 6}\,(\bar u u +\bar d d -2\bar s s)
\quad\mbox{or}\\
\quad \eta' &\approx&\frac{1}{\sqrt
3}\,(\bar u u +\bar d d + \bar s s)\,. \nonumber
\end{eqnarray}
This indicates a large violation of the Zweig rule in the $0^{-+}$
channel
\cite{fritzsch75} \cite{venez89}. Large transitions between the various
($\bar q q$)--configurations
must take place. In QCD the strong mixing effects are related to the
spontaneous breaking of the chiral $U(1)$ symmetry normally attributed
to
instantons. Effectively the mass term for the pseudoscalar mesons can
written as
follows, neglecting the effects of symmetry breaking in the gluonic
mixing term \cite{fritzsch73} \cite{hooft76} \cite{shifm86}:
\begin{equation}
\quad M^2_{\bar q q} = \left( \begin{array}{ccc}
M^2_u & 0 & 0\\ 0 & M^2_d & 0\\ 0 & 0 & M^2_s \end{array} \right)\, +\,\,
\lambda \left( \begin{array}{ccc} 1&1 &1\\ 1&1 &1\\ 1&1 &1\\            
 \end{array} \right)\,,
\end{equation}
where $M^{2}_{u},M^{2}_{d}$ and $M^{2}_{s}$ are the $M^{2}$--values
of the masses of quark composition $\bar u u, \bar d d$ and $\bar s s$
respectively.\\
It is well--known that the mass and mixing pattern of the
$0^{-+}$--mesons is described by such an ansatz \cite{fritzsch75}.
The parameter $\lambda$, which describes the mixing strength due to the
gluonic forces, is essentially given by the $\eta'$--mass:
$\lambda \cong 0.24$ GeV$^{2}$. Since $\lambda$ is large
compared to the strength of $SU(3)$ violation given by the $s$--quark
mass, large mixing phenomena are present in the $0^{-+}$ channel, as
seen in the corresponding wave functions.\\
The situation is different in the vector meson $1^{--}$ channel.
Here the gluonic mixing term is substantially smaller than the strength
of $SU(3)$ violation such that the Zweig rule is valid to a good
approximation. If one describe the mass matrix for the vector mesons
in a similar way as for the pseudoscalar, we have
\begin{equation}
M_{\bar q q} = \left( \begin{array}{ccc} M(\bar u u) & 0 & 0\\ 0 &
M(\bar d d) & 0\\ 0 & 0 & M(\bar s s) \end{array}\right)\,
+\,\,\tilde \lambda \left( \begin{array}{ccc} 1&1 &1\\
1&1 &1\\ 1&1 &1\\ \end{array} \right)\,,
\end{equation}
here $M(\bar q q)$ denotes the mass of a vector meson with quark
composition $\bar q q$ in the absence of the
mixing term. The magnitude of the mixing term $\tilde \lambda$ can be
obtained in a number of different ways, e.g by considering the
$\rho_{0}$--$\omega$ mass difference.
Neglecting the isospin violation caused by the $m_{d}$--$m_{u}$
mass splitting, the gluonic mixing term is responsible for the
$\rho_{0}$--$\omega$ mass shift:
\begin{equation}
\quad M_{\omega}-M_{\rho}= 2 \tilde \lambda\quad ,
\end{equation}
\begin{displaymath}
\quad\tilde \lambda \cong 6.0 \pm 0.5             
\,\,\mbox{MeV}\,.
\end{displaymath}
In QCD the isospin symmetry is violated by the mass splitting
between the $u$-- and $d$--quark. Typical estimates give:
\begin{equation}
\quad \frac{m_{d}-m_{u}}{\frac{1}{2}(m_{d}+m_{u})}\cong 0.58\,.
\end{equation}
The observed smallness of isospin breaking effects is usually attributed
to
the fact that the mass difference $m_{d}$ -- $m_{u}$ is small compared
to the QCD scale $\Lambda_{QCD}$. However in the case of the vector
mesons
the QCD interaction enters in two different ways:\\\\
a) \, In the chiral limit of vanishing quark masses the masses of the
vector mesons are solely due to the QCD interaction, i.e.
$M= \mbox{const}\cdot\Lambda_{QCD}$.\\\\
b)\, The QCD mixing term will lead to a mixing among the various flavour
components such that the $SU(3)$ singlet (quark composition $(\bar u u+
\bar d d +\bar s s)/\sqrt 3$) is lifted upwards compared to the two
other neutral components given by the wave functions $(\bar u u - \bar d
d)/\sqrt 2$ and $(\bar u u + \bar d d-2 \bar s s)/\sqrt 6$. The
corresponding mass shift is given by $3\tilde \lambda$.\\\\
We approach the real world by first introducing the mass of the strange
quark. As soon as $m_{s}$ becomes larger than $3\tilde \lambda$,
substantial
singlet--octet mixing sets in, and the mass of one vector meson
increases until it reaches the observed value of the $\phi$--mass.
At the same time the Zweig rule, which is strongly violated in the        
chiral
$SU(3)_{L}\times SU(3)_{R}$ limit becomes more and more valid.\\
The validity of the Zweig rule is determined by the ratio
$m_{s}/\tilde\lambda$. If this ratio vanishes, the Zweig rule is
violated strongly. In reality, taking $m_{s}$ (1GeV) $\approx$ 150 MeV,
the ratio $m_{s}/\tilde\lambda$ is about $25$ implying that the
Zweig rule is nearly exact.\\
In a second step we introduce the light quark masses $m_{u}$ and
$m_{d}$.
We concentrate on the non--strange vector mesons. If the gluonic
mixing interaction were turned off, the mass eigenstates would be
$v_{u}=|\bar u u\rangle$ and $v_{d}=|\bar d d\rangle$. The masses of
these mesons are given by:
\begin{equation}
\quad M(v_{u}) = \langle v_{u}|\, H^{0}+ m_{u}\,\bar u u\,|
\,v_{u}\rangle\,,
\end{equation}
\begin{displaymath}
\quad M(v_{d}) = \langle v_{d}|\, H^{0}+ m_{d}\,\bar d d\,|
\,v_{d}\rangle\,.
\end{displaymath}
Here $H^{0}$ is the QCD--Hamiltonian in the chiral limit
$m_{u}=m_{d}=0$. Thus the masses can be written as
\begin{equation}
\quad M(v_{u}) = M_{0}+ 2m_{u}\cdot c\,,
\end{equation}
\begin{displaymath}
\quad M(v_{d})= M_{0}+ 2m_{d}\cdot c\,.
\end{displaymath}
(c: constant, given by the expectation value of $\bar q q$).
The introduction of the light quark masses induces positive mass
shifts for both $v_{u}$ and $v_{d}$. These mass shifts can be estimated
by considering the corresponding mass shifts of the charged
$K^{*}$--mesons. One finds \cite{scadr84} \cite{fritmu}: 
\begin{eqnarray}
M(v_{d})-M(v_{u}) & \cong & 2\,(m_{d}-m_{u})\cdot c \\
& \cong & 1.7 \,\,\mbox{MeV}. \nonumber
\end{eqnarray}
It is remarkable that this mass shift is of similar order of magnitude
as the mass shift  between the isosinglet and isotriplet state in the
chiral limit, where isospin symmetry is valid. This implies that the
strength of the gluonic mixing term is comparable to the $\Delta I = 1$
mass term. If follows that the eigenstates of the mass operator taking
both the violation of isospin and the gluonic mixing into account will not
be close to being eigenstates of the isospin symmetry.\\
\\
For the $\rho_{0}$--$\omega$ system
the mass operator takes the form:
\begin{equation}
\quad M = \left( \begin{array}{cc} M(v_u) & 0 \\ 0 & M(v_d)
\end{array} \right)\, \,+\,\,\tilde \lambda
\left( \begin{array}{cc} 1&1 \\1&1 \end{array} \right)\,.
\end{equation}
Using $M(v_{u})= M(\bar u u), M(v_{d})= M(\bar d d)$ and
$\tilde \lambda= 5.9$ MeV, we find
\begin{eqnarray}
|\rho_{0}\rangle & = & 0.997 \Big | \frac{1}{\sqrt 2}(\bar u u -\bar d
d)\rangle + 0.071 \Big | \frac{1}{\sqrt 2}(\bar u u +\bar d
d)\rangle\nonumber \\
|\omega\rangle & = & -0.071 \Big | \frac{1}{\sqrt 2}(\bar u u -\bar d
d)\rangle + 0.997 \, \Big | \frac{1}{\sqrt 2}(\bar u u +\bar d
d)\rangle\nonumber
\end{eqnarray}                                                  
The mixing angle $\alpha$ discribing the strength of the
triplet--singlet mixing is about $-4.1^{o}$, i.e. a sizeable violation of
isospin symmetry is obtained. Neither is the $\rho_{0}$--meson an
isospin triplet, nor is the $\omega$--meson an isospin singlet.\\
\\
The conclusions we have derived follow directly from the observed
smallness of the gluonic mixing in the vector meson channel and the
$m_{u}-m_{d}$ mass splitting, as observed e.g. in the mass spectrum of the
$K^{*}$--mesons.
Nevertheless they are in direct conflict with observed facts. According
to eq. (13), the probability of the $\rho_{0}$--meson to be an
$I=|\frac{1}{\sqrt 2}(\bar u u +\bar d d)\rangle$--state is
$\mbox{sin}^{2}\alpha \cong 0.51\%$. Taking into account the observed
branching ratio for the decay $\omega \rightarrow
\pi^{+}\pi^{-}$, BR $\cong(2.21\pm 0.30)\%$, this probability is bound
to be less than $0.12\%$, in disagreement with the value derived above.
Obviously our theoretical estimate cannot be correct.\\
We consider the discrepancy described above as a serious challenge for
our understanding of the low energy sector of QCD.\\
We conclude: The mass difference $\Delta M = M(v_{d})-M(v_{u})$ must be
smaller than estimated above. In order to reproduce the observed branching
ratio for the decay $\omega \rightarrow \pi^{+}\pi^{-}$, $\Delta M$ cannot
exceed $0.82$ MeV, implying that our simple estimates based on
quark--model considerations, cannot be correct. This can be seen as
follows. We consider the following two--point functions
\begin{eqnarray}
u_{\mu\nu}&=& \langle 0|\bar u (x)\gamma_{\mu}u(x)\,
\bar u (y)\gamma_{\nu} u(y)|0\rangle\,, \nonumber \\
d_{\mu\nu}&=& \langle 0|\bar d (x)\gamma_{\mu}d(x)\,
\bar d (y)\gamma_{\nu} d(y)|0\rangle \\
m_{\mu\nu}&=& \langle 0|\bar d (x)\gamma_{\mu}d(x)\,
\bar u (y)\gamma_{\nu} u(y)|0\rangle\,. \nonumber
\end{eqnarray}
The mixed spectral function $m_{\mu\nu}$ is expected to be
essentially zero in the low energy region, since the two different
currents can communicate only via intermediate gluonic mesons. In
perturbative QCD these states would be represented by three gluons. The
vanishing of $m_{\mu\nu}$ implies the validity of the Zweig rule. \\
The spectral functions $u_{\mu\nu}$ and $d_{\mu\nu}$ are strongly
dominated
at low energies by the $\rho_{0}$-- and $\omega$--resonances. The actual
intermediate states contributing to the two-point functions are $2
\pi$--and $3 \pi$--states. However, a violation of the isospin symmetry
due to the $u-d$--quark mass splitting does not show up in the $\pi$-meson
spectrum. The $\pi^+- \pi^{\circ}$ mass splitting is solely due to the
electromagnetic interaction. It follows that resonant
$\left( 2 \pi \right)$ of $\left( 3 \pi \right)$ states, i. e. the
$\rho$--$\omega$--resonances, cannot display the effects of the isospin
violation either, and the mass difference
$\Delta M = M \left( v_d \right) - M \left( v_u \right)$
must be very small.\\
\\
Although the isospin symmetry is broken explicitly by the $u-d$ mass terms,
this symmetry violation does not show up in the $\rho$--$\omega$ sector.
The isospin symmetry breaking is shielded by the pion dynamics. Effectively
the symmetry is restored by dynamical effects. Here the gluon anomaly
plays an important role. The effect of a dynamical symmetry restoration
by nonperturbation effects discussed here might be reproduced by lattice
simulations. It might be that similar symmetry restoration effects are
present in other situations, for example in the electroweak sector, which
is sensitive to the dynamics in the TeV region.\\
\\
I am happy that this paper could be included in the volume dedicated to
my friend Hagen Kleinert on the occasion of his 60th birthday. We never
collaborated
together, but our wordlines met regularly e. g. at CERN, in Berlin, in
Pasadena. After introducing Hagen to Dick Feynman, I was happy to learn
that they finally wrote a paper together. I have problems seeing Hagen as
a sixty--years old colleague, since he appears and acts like a senior
post--doc. In Aspen Hagen would never be granted the price--reduction on
the lift--ticket, offered to anyone above 60, unless he takes his birth
certificate along -- but also in ten years from now Hagen will have the
same problem.

\end{document}